# Manifestation of ageing in the low temperature conductance of disordered insulators

Thierry Grenet and Julien Delahaye

*Institut Néel, CNRS & Université Joseph Fourier, BP 166, F-38042 Grenoble Cédex 9*

**Abstract** – We are interested in the out of equilibrium phenomena observed in the electrical conductance of disordered insulators at low temperature, which may be signatures of the electron coulomb glass state. The present work is devoted to the occurrence of ageing, a benchmark phenomenon for the glassy state. It is the fact that the dynamical properties of a glass depend on its age, i.e. on the time elapsed since it was quench-cooled. We first critically analyse previous studies on disordered insulators and question their interpretation in terms of ageing. We then present new measurements on insulating granular aluminium thin films which demonstrate that the dynamics is indeed age dependent. We also show that the results of different relaxation protocols are related by a superposition principle. The implications of our findings for the mechanism of the conductance slow relaxations are then discussed.

**Introduction** – Recently a set of out of equilibrium phenomena was reported and studied in the low temperature conductance of disordered insulators like granular metals [1, 2], indium oxide [3] and ultra thin films of metals [4]. Macroscopic relaxation times were observed and an electronic origin was suggested for them, which opens fascinating questions as processes involving current carriers are generally thought to be fast. In practice, after the systems studied are cooled from room to liquid helium temperature, their conductance logarithmically decreases for days or weeks without any sign of saturation. If inserted in MOSFETs as their channel (see Fig 1 a), they exhibit an anomalous field effect which consists of a symmetrical minimum (thereafter called «the dip») in the conductance versus gate voltage ($G(V_g)$) curves (upper curve in Fig. 1 c). The logarithmic decrease of G *at fixed* $V_g$ observed after cooling is associated with the progressive formation of a dip centred on that $V_g$ value. The properties and slow dynamics of the conductance dip have been studied in some detail in indium oxide [5, 6, 7, 8] and in granular aluminium [9].

An interpretation envisaged for these relaxation phenomena [5] rests on the concept of an electronic coulomb glass, a phase the carriers could constitute owing to their localized and interacting character [10]. One expects that the electronic coulomb glass possesses generic glassy dynamical features like ageing [11]. As a glass is out of equilibrium and endlessly evolves towards an equilibrium state, its microscopic state is always changing and one expects that most of its properties evolve with time. In this paper we use the term ageing with the following meaning: ageing is the fact that during that never ending evolution the *dynamical* properties of the glass evolve [12], it generally responds more and more slowly to external stimuli. Several previous experiments have been interpreted as evidences of ageing in indium oxide [13, 14, 15, 6, 7] and by ourselves in granular aluminium [9].

In the present paper we first recall and critically discuss previous experiments and their interpretation. In particular we show that they cannot be taken as demonstrations of ageing according to the aforementioned definition of this term, as strictly speaking no unambiguous change of the *dynamics* was revealed. We then present new relaxation measurements on granular aluminium films using modified protocols, which demonstrate qualitatively new phenomena which we interpret as a clear signature of ageing. We explain why these could not be observed in previous experiments, and we show how the relaxations in different experimental protocols can be related to each other by the application of a superposition principle. We finally discuss the implications of our work for the "electronic coulomb glass" and other competing interpretations of the conductance dip and its slow relaxation in disordered insulators.

**Critical analysis of previous "ageing" experiments** – To illustrate our discussion, we recall briefly the previous experiments on insulating granular Al thin films, similar to the ones also performed on indium oxide. We are interested in the slow response of the conductance to sudden gate voltage changes at low temperature. To look for ageing effects the following protocol was generally used (see Fig.1 b):

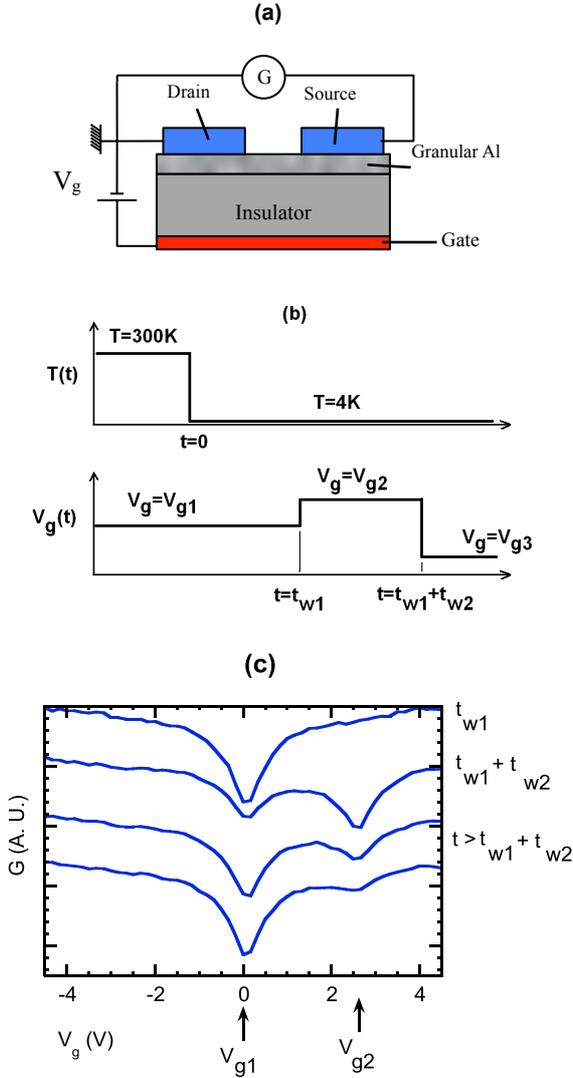

Fig. 1: (a) Scheme of a MOSFET like device used to measure the electric field effect in granular aluminium. (b) Scheme of the temperature and gate voltage sequences used in the "three step protocol". (c) Illustration of the field effect anomalies and their dynamics. After the sample has been cooled to 4K with $V_g = V_{g1}$, a dip forms during $t_{w1}$, centred on $V_{g1}$ (upper curve). If $V_g$ is then set to $V_{g2}$ (2.5V in this illustration) a new dip forms while the first one vanishes (second upper curve at $t_{w1}+t_{w2}$). If $V_g$ is set back to $V_{g1}$ the first dip is restored while the second one is progressively erased. We are mainly interested in the dynamics of the dip at $V_{g2}$. The curves are shifted for clarity.

- First step: after the sample has been rapidly cooled from room temperature to $T_1$=4K (with $V_{g1}$=0), a dip (we call dip 1) progressively forms in the $G(V_g)$ curve and can be visualized by performing a fast $V_g$ scan like in the upper curve of Fig. 1c. As the sample slowly proceeds towards equilibrium, the dip 1 amplitude increases like Ln($t$) ($t$ being the time since the fast cool).
- Second step: at time $t_{w1}$, $V_g$ is switched to $V_{g2}$ and the formation of a second dip (dip 2) occurs while the dip 1 progressively vanishes (second upper curve of Fig. 1c).
- Third step: at time $t_{w1}+t_{w2}$, the gate voltage is set to a third value $V_{g3}$, the dip 2 is consequently progressively erased (two lower curves of Fig. 1c, where incidentally in this illustration $V_{g3} = V_{g1}$).

We call this protocol the "three steps protocol". It is similar to the isothermal-remanent magnetization (IRM) experiment used in the context of spin glasses (see for example [16]).

An important point is that these experiments were performed after the sample had long "equilibrated" with $V_g = V_{g1}$, reaching a "quasi-equilibrium state" (i.e. $t_{w1} >> t_{w2}$). It was found that the writing of the dip 2 is history free (*new dip amplitude increasing like Ln(t)*), while the dip 2 was erased as shown in Fig. 2. The erasure curves clearly depend on $t_{w2}$ : the longer $t_{w2}$ is, the longer it takes to erase the dip 2. This was interpreted as ageing. Moreover it is seen that the erasure curves all precisely collapse on an "erasure master curve" when plotted versus $t/t_{w2}$ ("full scaling"), and the characteristic erasure time (defined as the intercept between the extrapolated logarithmic part and the abscissa axis, see inset of Fig.2) is equal to $t_{w2}$. The same observations were reported with indium oxide (see e.g. Figs 11 and 12(a) of [14])[1] . This dependence on $t_{w2}$ of the relaxation back to the quasi-equilibrium state was taken as a signature of "full ageing" (ageing with full scaling).

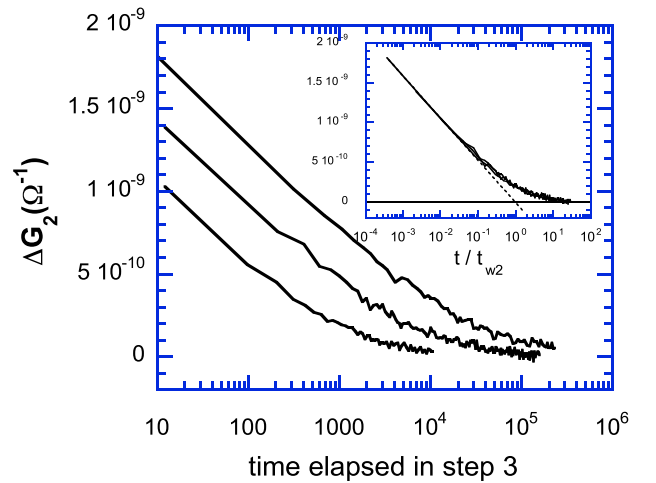

Fig. 2: Time dependence and full scaling of the vanishing amplitude of the dip 2 during the third step of the three step protocol (see text). Main graph: dip 2 amplitudes for $t_{w2}$ = 1080 sec, 5400 sec, 27000 sec from left to right. The inset shows the perfect collapse of the curves plotted as a function of the reduced time $t/t_{w2}$ (from [9]).

We now discuss these previous studies and point to their shortcomings. We first note that an actually more natural and simpler ageing protocol would be to restrict ourselves to the first and second steps of the three step protocol. Indeed if the dynamics of the systems does depend on their age, then the growth of the dip 2 during $t_{w2}$ should depend on the value of $t_{w1}$. In particular if the system becomes "stiffer" due to the growth of some internal correlations (the usual qualitative interpretation of ageing) one should observe that the writing of a new dip is slower when the system has spent a longer time in its glassy state. Thus the amplitude of the dip 2 written for a given $t_{w2}$ should be smaller for longer $t_{w1}$. Unfortunately no such effect was reported yet nor was observed by us previously, which questions the actual occurrence of ageing. One similar experiment was made with indium oxide in [7] (there called "T protocol"). Unfortunately the question of the $t_{w1}$ dependence of the growth of the dip 2 was not discussed. Instead a logarithmic extrapolation of the $t_{w1}$ relaxation curves was subtracted from the raw dip 2 growth curves. The resulting curves then of course strongly depend on $t_{w1}$, but the justification of this procedure is not clear to us and was not given in [7]. Actually we will show in the discussion section of the present paper how ageing (i.e. $t_{w1}$ dependence of the growth of the dip 2) may be discerned in the subtracted curves shown in [7].

---

[1] To be precise in indium oxide the master function was measured by following the return of the dip 1 to its "equilibrated" amplitude after $t_{w2}$. Since as long as $t_{w1} >> t_{w2}$ (and in indium oxide as long as the $V_g$ change is not too strong) the time evolutions of dips 1 and 2 are symmetrical [15, 9], the same master function as ours was obtained.

Moreover the $t_{w2}$ dependence and "full scaling" observed with the three step protocol are not an undisputable evidence of ageing. One way to show this is to recall that one can reproduce the curves of Fig. 2 and their full scaling using a simple phenomenological model which does not incorporate ageing, as we have already shown and only briefly recall here. For more details see [9]. We supposed that a collection of reversible slow "degrees of freedom" influence the percolating conducting channels in the samples and that their relaxation "back" and "forth" can be induced by gate voltage changes. The precise nature of such degrees of freedom is not important for our argument. One may for instance think of localized polarisable tunnelling systems as has been considered in [9, 17]. Since the conductance relaxations observed are rather small (a few percents) we admit that the effects of the individual degrees of freedom are small and additive. Their relaxation times $\tau_i$ are exponentially dependent on the broadly distributed parameters of the tunnelling barriers, which results in a $1/\tau_i$ distribution of relaxation times. With these simple ingredients one obtains the logarithmic growth of a new cusp and we showed in [9] that one also reproduces quite accurately the measured erasure curves and their full scaling without any fitting parameter. But this model does not incorporate ageing: the $\tau_i$ do not depend on the system's age. Actually the slopes of the logarithmic parts in Fig. 2 are all the same, which suggests that the dynamics of the erasure is the same in all cases and does not depend on $t_{w2}$. Obviously the difference between the curves simply comes from the fact that they do not start from the same conductance value. The intercept of the logarithmic part of the erasure curves define typical erasure times that are equal to $t_{w2}$. In other words it takes the same time to erase the dip 2 as it took to write it, which sounds quite natural.

Thus as the experimental results discussed so far do not show any age dependence of the relaxation times at work, and can be reproduced using a simple model which does not incorporate ageing in that meaning, we conclude that they are not a demonstration of ageing.

Note for completeness that the $1/\tau_i$ distribution is also expected for the correlated coulomb glass model as was anticipated in [18] and was recently justified more quantitatively in [19], where the erasure phenomenology is also derived (to our point of view improperly referred to as "ageing").

We now turn to our experimental program looking for ageing. It is simple in principle: systematically look for some effect of $t_{w1}$ on the *dynamics* of the dip 2 writing during $t_{w2}$ ("two step protocol"), and if any is found, look for the corresponding effect on the *dynamics* of the dip 2 erasure ("three step protocol"). We have of course no reason to restrict to $t_{w1} \gg t_{w2}$ as was done in the previous experiments. We shall now describe the results of such a study which we performed on granular aluminium films.

**Experimental manifestations of ageing** – The results below were obtained studying two different samples which consist of field effect devices having a 200Å thick channel made of insulating granular aluminium. Sample A was prepared on a sapphire substrate on which an Al gate and a 1000Å thick alumina gate insulator were deposited prior to the channel. Sample B was deposited on a doped silicon wafer, the 1000Å thick thermally grown oxide layer acting as the gate insulator. The granular aluminium films had $R_\square(4K)$ = 20 GΩ for sample A and $R_\square(4K)$ = 490 MΩ for sample B. The relaxation experiments were performed at liquid He temperature unless specified, the conductance was measured using either an AC or a DC voltage excitation and a current to voltage amplifier for current measurements. Details about the samples, measurement techniques and precautions were already given in [2, 9]. Similar results were systematically obtained for both samples A and B.

To follow the growth dynamics of the dip 2 in the two step protocol, we measured its amplitude evolution $\Delta G_2(t) = G(V_{gref}) - G(V_{g2},t)$ where $t$ is the time spent in step 2. The base line conductance $G(V_{gref})$ was measured at each time $t$ in order to eliminate any effect of possible He bath temperature drifts which could shift all the $G(V_{gi})$ by some small time dependent value. The excursions of $V_g$ out of $V_{g2}$ were kept short so that the growth of dip 2 is not perturbed and a prominent dip 3 is not formed at $V_{gref}$ (the system spent more than 95% of the time with $V_g = V_{g2}$). We found that, while for times much smaller than $t_{w1}$ the dip 2 grows logarithmically as expected, in the region $t \approx t_{w1}$ the growth curves depart from the pure Ln($t$): the dip growth is somewhat faster around $t_{w1}$ and seems to decelerate for $t \gg t_{w1}$. This is seen in Fig. 3-a (sample A) and in Fig. 4-a (sample B) where we show growth curves $\Delta G_2(t)$ obtained for different $t_{w1}$s spanning over several orders of magnitudes. The departures from the pure Ln($t$) growth are clearly seen in Fig. 3-b and in Fig. 4-b where we show the differences from the pure logarithmic relaxations, plotted versus $t$ and $t/t_{w1}$. The pure Ln($t$) growth are determined for each sample by the logarithmic part of the highest $t_{w1}$ curve. It is seen that the deviations from Ln($t$) roughly scale with the time spent at $V_{g1}$. Another way to visualize the $t_{w1}$ dependence is to look at the derivatives of the relaxation curves. This could only be performed after a Fourier transform filtering of the noise. It is seen in Figs. 3-c and 4-c that the derivatives exhibit a broad maximum which corresponds to the inflexion of the relaxation curves (in logarithmic scale) and which position also scales with $t_{w1}$.

The departures from Ln($t$) amount to roughly 10% of the observed relaxations. In spite of their limited amplitude, they constitute a qualitatively new phenomenon and show that the growth dynamics of dip 2 indeed depends on $t_{w1}$, a clear signature of ageing according to our definition of it.

This behaviour is reminiscent of e.g. ageing effects seen in zero field cooled magnetization relaxation experiments of spin glasses [20]. In this case the evolution of the relaxation curves was interpreted as the signature of an age dependent relaxation time distribution. The latter is usually computed as the derivative of the relaxation curves versus Ln($t$), and exhibits a broad maximum centred on the samples age like in our case.

We used another procedure to demonstrate the age effect: measure, for different $t_{w1}$ values, the dip 2 amplitude obtained for a fixed duration $t_{w2}$ spent with $V_g = V_{g2}$. Practically the dip 2 amplitude was measured by switching the gate voltage to a value $V_{gref}$ just after $t_{w2}$ and measuring the conductance jump induced ($\Delta G_2$), which is simply the difference between the reference base line conductance (immediately measured at $V_{gref}$) and the bottom of the dip 2 (see Fig 5). The $V_{g2}$ and $V_{gref}$ values were +5V and -5V respectively for sample A, and +10V and -10V for sample B. The differential procedure ensures that the determination of $\Delta G_2$ is not significantly perturbed by possible slow drifts of the He bath and sample temperature[2].

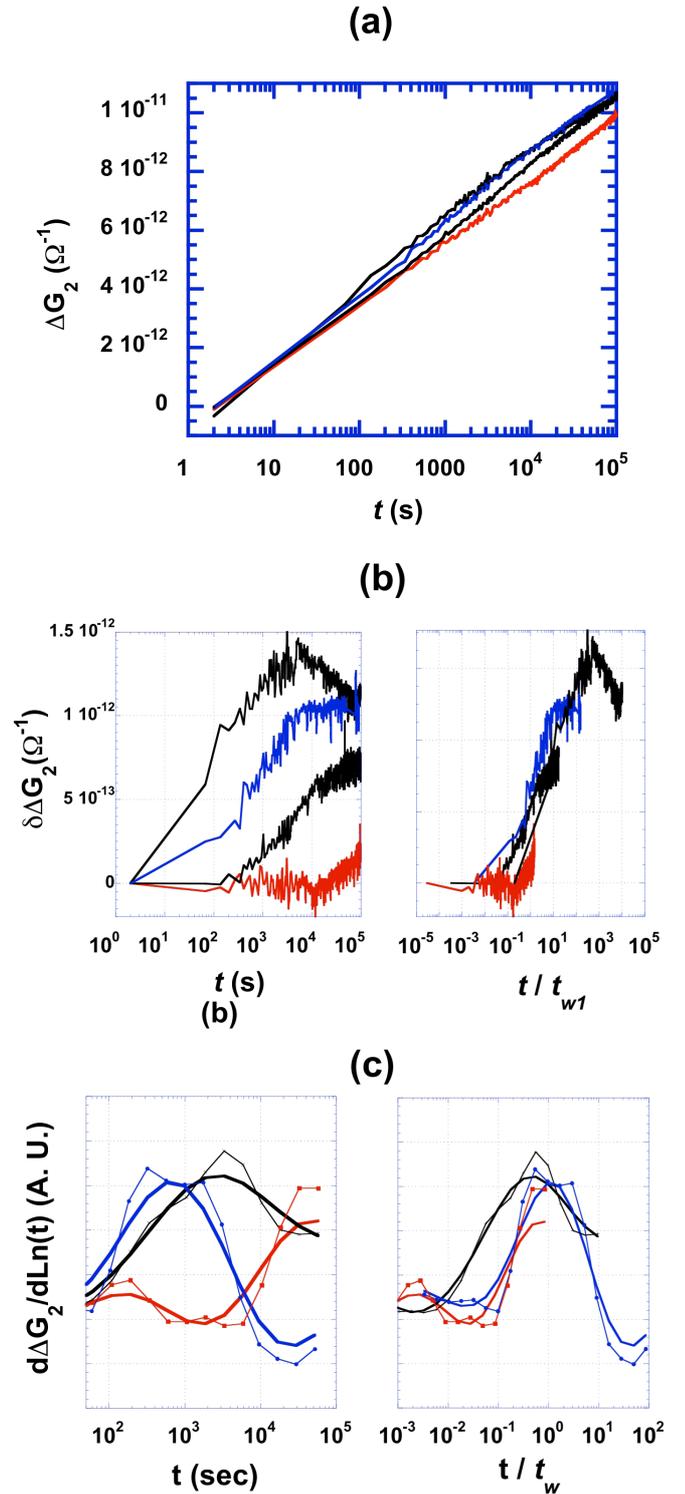

Fig. 3: Ageing effect in the growth of the dip 2 in sample A. (a): dip 2 amplitude as a function of time for different $t_{w1}$ values, from bottom to top curves: $t_{w1}$ = 6.85 10$^4$ s, 6000 s, 600 s and 10 s; (b): departures from the pure Ln($t$) growth as a function of $t$ (left) and the reduced time $t/t_{w1}$ (right); (c): derivative of $\Delta G_2$ as a function of $t$ (left) and the reduced time $t/t_{w1}$ (right). The curves are shown for two filterings to show that the broad maximum position does not depend on the amount of filtering.

---

[2] We estimated that with this procedure the temperature drifts sometimes observed (a few mK during an experiment) could not influence the $\Delta G_2$ relaxations by more than one tenth of the effects we discuss in this paper.

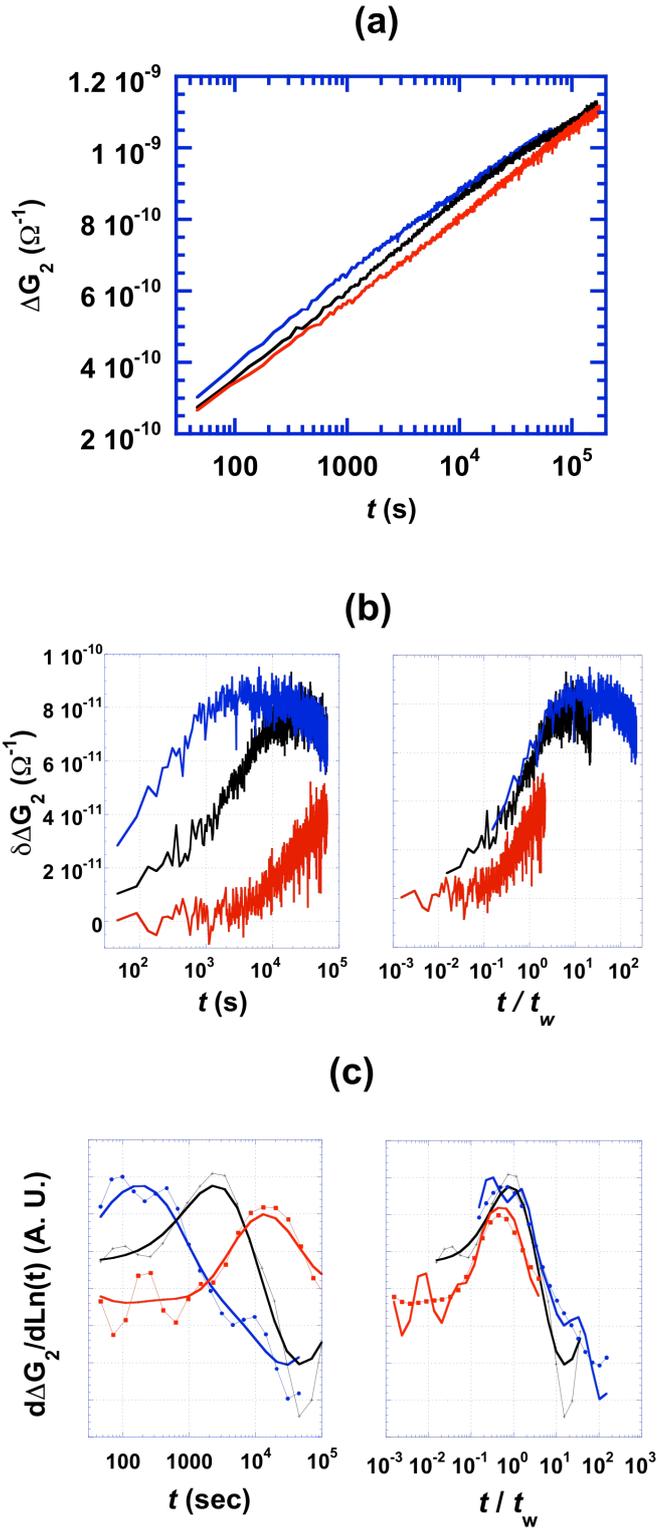

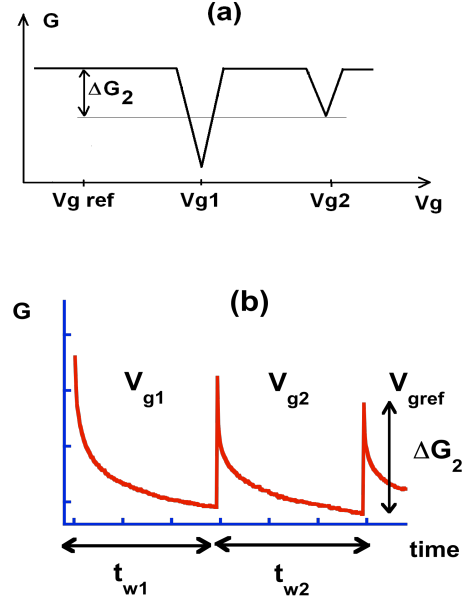

Fig. 4: Ageing effect in the growth of the dip 2 in sample B. (a): dip 2 amplitude as a function of time for different $t_{w1}$ values, from bottom to top curves: $t_{w1}$ = 3 $10^4$ s, 3000 s and 300 s; (b): departures from the pure Ln(t) growth as a function of $t$ (left) and the reduced time $t/t_{w1}$ (right); (c): derivative of $\Delta G_2$ as a function of $t$ (left) and the reduced time $t/t_{w1}$ (right).

This measurement was repeated for different $t_{w1}$ (and the same $t_{w2}$). Between each measurement the sample was heated up to 90K so as to erase any dip, the fast (not more than 10 seconds) cool down from 90K to 4K was realized by plunging it in the liquid He. The samples proved to be very stable upon this thermal cycling.

Fig. 5: Procedure used to investigate the effect of $t_{w1}$ on the amplitude of the dip 2 written for a given $t_{w2}$. The sample is first quenched to 4K and kept during $t_{w1}$ with $V_g = V_{g1}$ (step 1). Then $V_g$ is switched to $V_{g2}$ and the dip 2 is formed during $t_{w2}$ (step 2). Finally $V_g$ is set to $V_{gref}$ and the amplitude of the dip 2 is determined as the conductance jump $\Delta G_2$ induced by this last $V_g$ change. In (a) we schematize the $G(V_g)$ curve which would be obtained if a fast $V_g$ scan was performed after the step 2 and indicate $\Delta G_2$. In (b) we show the actual conductance variations measured and indicate $\Delta G_2$.

In Fig. 6 we show the results of measurements performed with sample A. Here $t_{w2}$ = 1350 s and $t_{w1}$ was varied over more than two orders of magnitude. One observes the $\Delta G_2$ dependence on $t_{w1}$. As expected from Fig. 3 the longer the sample was first aged at 4K with $V_g = V_{g1}$, the smaller the achieved dip 2. The straight line in the figure is only a guide to the eyes, as one can expect that the effect saturates for large $t_{w1}$ values [3].

One may wonder what perturbation applied to the system after $t_{w1}$ may rejuvenate it. Obviously, a change in gate voltage $V_g$ of a few volts does not have such an effect otherwise the ageing would not be observable in our experiments (the age of the system would be erased when $V_g$ is switched from $V_{g1}$ to $V_{g2}$). This can be further demonstrated in the following experiment: the system was aged for $t_{w1a}$ = 13500 s (i.e. $10 t_{w2}$) at $V_{g1a}$ = -2,5V and then for $t_{w1b}$ = 2250 s (i.e. 1,67$t_{w2}$) at $V_{g1b}$ = +2,5V (for the subsequent steps $V_{g2}$ = 7,5V and $V_{gref}$ = -7,5V). It is seen in

---
[3] Note that to describe the data, we mention $t_{w1}/t_{w2}$ values in order to specify the situations regarding the $t_{w1} \ll t_{w2}$ and $t_{w1} \gg t_{w2}$ limits. But we do not know whether the precise behaviours are uniquely determined by the $t_{w1}/t_{w2}$ ratios for any $(t_{w1}, t_{w2})$ as we have not yet repeated the experiments for different $t_{w2}$ values.

Fig. 6 (filled triangle) that the $\Delta G_2$ value thus measured corresponds well to the value expected for $t_{w1} \approx 10 t_{w2}$ and definitely not to $t_{w1} \approx 1{,}67 t_{w2}$. Hence the gate voltage change from $V_{g1a}$ to $V_{g1b}$ has not rejuvenated the sample, which age is of the order of the total $t_{w1a} + t_{w1b}$. We do not know yet whether much larger $V_g$ steps can change the sample's age.

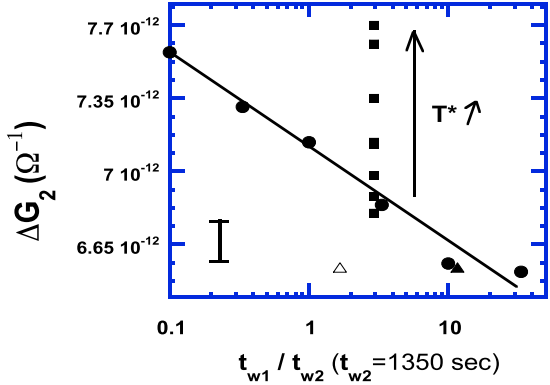

Fig 6: Effect of $t_{w1}$ on the amplitude of the dip 2 formed during $t_{w2}$ = 1350 s, for sample A. All steps performed at 4K unless specified. Filled circles show that the longer the sample was aged with $V_g = V_{g1}$, the smaller the dip 2 amplitude. The filled triangle corresponds to the sample aged at two different values of $V_{g1}$ (see the text), and shows that the two ageing steps are cumulative (if not the experimental point would instead be the empty triangle). Filled squares correspond to the cases where an excursion to a higher temperature T* was imposed between $0.75 t_{w1}$ and $t_{w1}$. The higher T*, the more the ageing at 4K between $t=0$ and $0.75 t_{w1}$ is erased. The T* values are: 6K, 8K, 12.5K, 16.5K, 20K, 26K, 36K and 46K. Error bars (shown on the left) are of the order of $2.10^{-13}\ \Omega^{-1}$.

One may expect that the effective age of the system is modified by heating it. Such an effect is shown in Fig. 6. For this series of measurements a "high temperature" excursion was imposed during the ageing step ($t_{w1}$). More precisely the latter was performed in two steps, the first one for $t_{w1a}=3960$ sec (i.e. $2{,}9 t_{w2}$) at the usual temperature $T=4K$ and the second one for $t_{w1b} \approx 1200$ sec (i.e. $0{,}9 t_{w2}$) at a higher temperature which we call T*. The sample was then rapidly cooled back to $T=4K$ and the gate voltage set to $V_{g2}$, the rest of the procedure being as usual. A series of measurements was performed with T* ranging from 4K to 46K. The corresponding points are filled squares in Fig. 6: the higher T*, the larger $\Delta G_2$. We naturally define an effective age of the sample as the ageing time which would have given the same $\Delta G_2$ if the whole process had been performed at $T=4K$. In Fig. 7 one sees that the effective age of the system is approximately exponentially reduced by the high temperature T*.

We finally turn our attention to the "erasure master curve" which was under scrutiny in previously published three step experiments. We want to know whether it is also influenced by $t_{w1}$. Recall we define the characteristic erasure time at the intercept between the extrapolated logarithmic part of the $\Delta G_2(t)$ and the time axis. As we already mentioned above when $t_{w1}$ is much larger than $t_{w2}$, this erasure time is found to be equal to $t_{w2}$. However when $t_{w1}$ is not very large, we find that the erasure master function is modified: the full scaling is still obtained within our experimental accuracy but the erasure time is larger than $t_{w2}$. This is shown in Fig. 8. Thus for a "young" system it takes a longer time to erase a dip than it took to write it. This modification of the erasure master curve is the actual signature of ageing in the three step protocol.

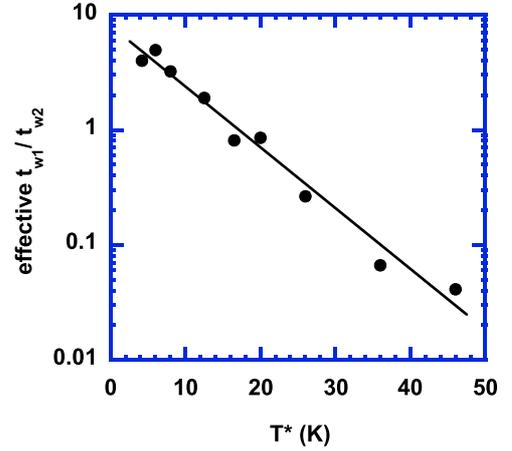

Fig. 7: Effective ageing time (in units of $t_{w2}$) as a function of T* for sample A. The effective $t_{w1}$ is defined as the ageing time which would give the same $\Delta G_2$ would the ageing be performed at the constant temperature $T=4K$. The effective $t_{w1}$ is exponentially reduced by the excursion at temperature T*.

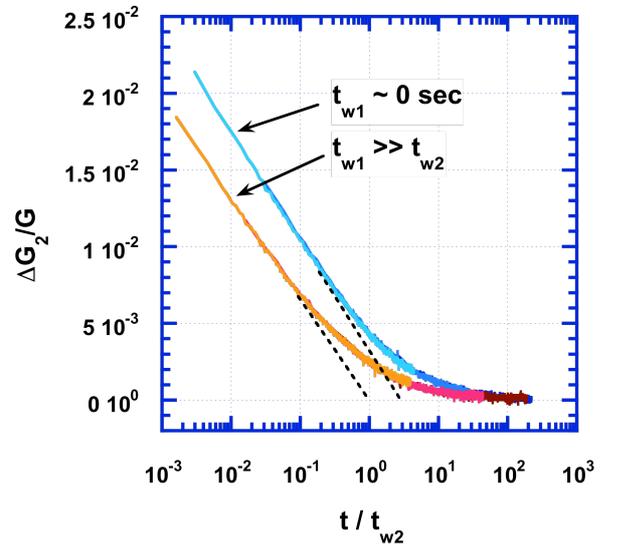

Fig. 8: Erasure of the dip 2 as a function of $t$ (time spent in step 3), when the gate voltage has been set back to $V_{g1}$ (sample B). The master curves are obtained over more than 5 orders of magnitude of $t/t_{w2}$ by superposing parts of it obtained with $t_{w2}$ = 300 s, 3000 s and 30000 s (the $t/t_{w2}$ spans of the superposed parts are: 0.002→4, 0.02→40, 0.2→200). Upper curve: the gate voltage was set to $V_{g2}$ upon quenching (i.e. $t_{w1}$ = 0 s). Lower curve: $t_{w1}$ was at least 15 times larger than the $t_{w2}$ values. The erasure relaxation functions are not the same for a "young" (upper curve) and an "old" (lower curve) sample. For a "young" sample, the curve starts from a more prominent dip 2 and the characteristic erasure time is larger than $t_{w2}$.

One may expect that the relaxations studied in the two step and three step protocols are related. We now show that the "erasure master curves" of Fig. 8 can indeed be deduced from the growth curves of Figs. 3 and 4 using a superposition principle analogous to the one found in spin glasses [16]. For clarity $\Delta G_2$ will be denoted $\Delta G_{growth}$ when the dip 2 is being written and $\Delta G_{erasure}$ when the dip 2 is being erased. Let then $\Delta G_{growth}(t, t_w)$ denote the growth with time $t$ of a new dip which writing started after the sample spent a duration $t_w$ at low temperature. The amplitude of a dip centred at some given $V_{g0}$ increases as long as $V_g$ is kept at $V_{g0}$, but it decreases if $V_g$ is set at some other value out of the $V_{g0}$ dip, as if a contribution of the opposite sign was added. We may then expect the erasure curve of the three step protocol to obey the following relation (choosing the origin of the time scale $t$ at the beginning of the third step):

$$\Delta G_{erasure}(t, t_{w1}, t_{w2}) = \Delta G_{growth}(t + t_{w2}, t_{w1}) \\ - \Delta G_{growth}(t, t_{w1} + t_{w2}) \quad (1)$$

This procedure is illustrated in Fig. 9-a. We show in Fig. 9-b two erasure curves computed from experimental growth curves of sample A (two step protocol) using Eq.1, in the two limiting cases $t_{w1} \gg t_{w2}$ and $t_{w1} \ll t_{w2}$. Note that in the first case $\Delta G_{growth}$ doesn't depend on $t_{w1}$ and is logarithmic, so that eq. (1) gives:

$$\Delta G_{erasure}(t, t_{w1}, t_{w2}) = A Ln(1 + \frac{t_{w2}}{t}) \quad (2)$$

This is just the erasure master curve predicted by the models mentioned in the previous section and shown to reproduce quite well the curves of Fig. 2 [9, 19]. However in the second case the $\Delta G_{growth}$ are not exactly logarithmic and the erasure curve computed from Eq. (1) is different. It is found to be similar to the measured erasure curve in Fig. 8. In particular the variation of the typical erasure time (extrapolation of the logarithmic part) with $t_{w1}$ is well reproduced. It is probable that using the superposition principle one can predict the relaxation following any gate voltage protocol once the experimental set of growth curves $\Delta G_{growth}(t, t_w)$ has been measured for a variety of $t_w$ values.

**Discussion** – The experiments described above show that the slow dynamics of the conductance dips depends on the system's age. This is revealed by variations of the dip "growth" and "erasure" curves when "young" and "old" systems are compared. This could not be observed in previous experiments in granular Al [2, 9] and in indium oxide [13, 14, 15, 6, 7] which only focused on the erasure curve for very old samples ($t_{w1} \gg t_{w2}$).

It would be interesting to know whether the ageing phenomena reported here also exist in indium oxide. One can find in [7] an indication that it may be so. Indeed if the $t_{w2}$ relaxation was exactly logarithmic in that case, the subtraction of the logarithmically extrapolated $t_{w1}$ relaxation should result in $\Delta G \propto Ln(1+t_{w1}/t)$ (apart from the vertical shift due to the normal field effect observed in InO$_x$). The logarithmic part of the resulting curve plotted versus $t/t_{w1}$ should then extrapolate to 1 like in our lower curve in Fig. 8. However an inspection of the Fig. 2-b in [7] shows that, in spite of the scatter of the data, the scaled curves rather seem to extrapolate to $t/t_{w1}$ larger than one. This means that the $t_{w2}$ relaxation in indium oxide also has a departure from pure $Ln(t)$ similar to ours. Note that according to our superposition principle the subtraction performed in the "T protocol" of [7] is actually a way to mimic the erasure curve in a three step protocol in the limit $t_{w1} \ll t_{w2}$ and is thus expected to extrapolate to a value larger than one. Note also that if we perform the same subtraction of $Ln(t)$ to our $t_{w2}$ relaxation curves we of course get similar results.

Interestingly it was reported in indium oxide that when one applies $V_g$ jumps [15] or non ohmic bias voltages [6] large enough to erase previously formed dips, then modified erasure master curves are obtained. For moderately large perturbations, they are similar to the upper one in Fig. 8 and exhibit full scaling and an increased characteristic erasure time. This is fully consistent with our results and shows that such modified erasure curves are obtained either with "trully young" systems (this work) or with "old" ones which have been rejuvenated by applying a high enough electrical disturbance. The systems are effectively "young" when there is no well formed dip at any gate voltage.

What can we learn from these ageing effects about the microscopic origin of the dip and its slow relaxation? Two interpretations have been previously envisaged to explain it. In the first, as we recalled in the introduction of this paper, it is supposed that the charge carriers form a coulomb glass phase, and the field effect dip reflects the relaxation to equilibrium, a suggestion that has triggered numerous theoretical studies [19, 21, 11, 22]. These consider model systems where electrons are strongly localized and their coulomb repulsion is unscreened. The dip formation at constant $V_g$ would then reflect the slow relaxation of the glass towards its ground state for this $V_g$. In the case of granular Al, the electrons are localized on the nanometric metal islands, and the slow relaxation of the electron glass would correspond to redistributions of electrons between the islands via correlated inter-island hopping.

In the context of granular metals, an "extrinsic" interpretation was also discussed [1, 23, 9]. The slow relaxation of conductance would possibly not be intrinsic to the charge carriers but be provoked by the slow relaxation of the disorder potential, which may be of atomic or ionic origin. It was shown that the slow relaxation of the dielectric polarization around the metallic islands (e.g. due to two level systems switching) could influence the coulomb blockade in such a way that the macroscopic conductance is reduced, thus creating a dip. A related model based on two level fluctuators was also suggested for the case of indium oxide [17]. However in the latter case, the effect of electron doping by oxygen deficiency is consistent with the electron coulomb glass scenario [24].

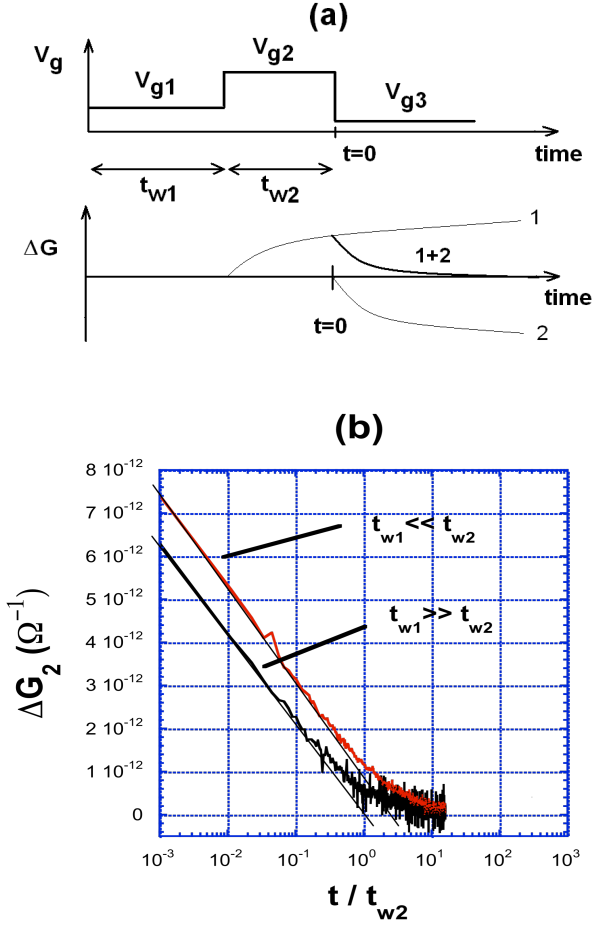

Fig 9: a – Scheme of the application of the superposition principle. Upper graph: $V_g$ sequence of the three step protocol. Lower graph: contributions to the amplitude of the dip situated at $V_g=V_{g2}$: "1" is $\Delta G_{growth}(t+t_{w2}, t_{w1})$, "2" is $-\Delta G_{growth}(t, t_{w1} + t_{w2})$ (see text). b - dip 2 amplitude as a function of the reduced time $t/t_{w2}$ *computed* using the superposition principle and the measured curves of Fig. 3 ($t_{w2}$ = 6000 sec, $t_{w1}$ = 10 sec and $6.85\ 10^4$ sec). Note the similarity with Fig. 8.

In [9] we noted that not only the extrinsic scenario, but also the coulomb glass one, can probably be described in terms of a collection of slow fluctuators influencing percolating conduction paths. This is suggested by the fact that the slow relaxation effects are observable in films close to the metallic state (with $R_\square(4K)\sim 100k\Omega$). While it is difficult to accept that macroscopically slow electrons can produce such a low resistance, one can imagine that due to inhomogeneities, almost metallic conducting path can be influenced by slow fluctuators composed of nearby highly insulating areas of electron glass. This has been considered more quantitatively in the case of indium oxide [22].

Our present experimental results indicate that the fluctuators cannot be simple independent reversible objects, but must be more complex or correlated objects which dynamical properties evolve with their history. This is consistent with the coulomb glass scenario and puts new constraints on the "extrinsic" models. Their simplest versions (effect of isolated two level systems) are ruled out, but one cannot exclude more sophisticated extrinsic mechanisms [25].

**Conclusions** – We have presented a study of the off equilibrium phenomena observed in the low temperature conductance of disordered insulators. We have shown how the slow dynamics observed in relaxation experiments depends on the systems' age, i.e. on the time elapsed since they were quench cooled, and that the responses to different gate voltage stimulations are related by a superposition principle. The ageing phenomenon is a bench-mark for glasses and its observation is indicative of a glassy state of Anderson insulators at low temperature.

\* \* \*

We acknowledge financial support from the "Jeunes Chercheurs" program of the French National Research Agency ANR (contract n° JC05_44044), as well as discussions with A. Amir who pointed out to us the reference [16].